\documentclass[journal=jacsat,manuscript=article]{achemso}

\usepackage[version=3]{mhchem} 
\usepackage{xcolor}
\usepackage{ulem}

\newcommand{\red}[1]{{\textcolor{black}{#1}}}

\author{Jinyoung Kim}
\affiliation{Department of Physics and Astronomy, Seoul National University, Seoul, 08826, Korea}
\author{Minjae Kim}
\email{garix.minjae.kim@gmail.com}
\affiliation{Department of Semiconductor Science and Technology, Jeonbuk National University, Jeonju, 54896, Korea}
\alsoaffiliation{Korea Institute for Advanced Study, Seoul, 02455, Korea}

\author{Donghan Kim}
\affiliation{Department of Physics and Astronomy, Seoul National University, Seoul, 08826, Korea}

\author{Sungsoo Hahn}
\affiliation{Department of Physics and Astronomy, Seoul National University, Seoul, 08826, Korea}

\author{Younsik Kim}
\affiliation{Department of Physics and Astronomy, Seoul National University, Seoul, 08826, Korea}

\author{Minsoo Kim}
\affiliation{Department of Physics and Astronomy, Seoul National University, Seoul, 08826, Korea}

\author{Byungmin Sohn}
\email{bsohn@skku.edu}
\affiliation{Department of Physics, Sungkyunkwan University, Suwon, 16419, Korea}

\author{Changyoung Kim}
\email{changyoung@snu.ac.kr}
\affiliation{Department of Physics and Astronomy, Seoul National University, Seoul, 08826, Korea}
\title[An \textsf{achemso} demo]
  {Low-dimensionality-induced tunable ferromagnetism in SrRuO$_3$ ultrathin films}

\begin{document}

\begin{tocentry}
\includegraphics{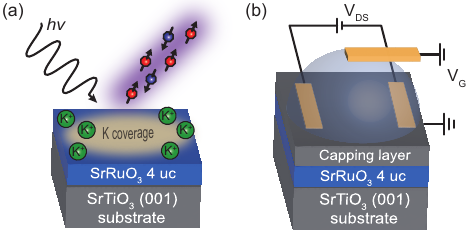}

\end{tocentry}

\begin{abstract}
Quantum materials near electronic or magnetic phase boundaries exhibit enhanced tunability, as their emergent properties become highly sensitive to external perturbations. Here, we demonstrate precise control of ferromagnetism in a SrRuO$_3$ ultrathin film, where a high density of states (DOS), arising from low-dimensional quantum states, places the system at the crossover between a non-magnetic and bulk ferromagnetic state. Using spin- and angle-resolved photoemission spectroscopy (SRPES/ARPES), transport measurements, and theoretical calculations, we systematically tune the Fermi level via electron doping across the high-DOS point. We directly visualize the spin-split band structure and reveal its influence on both magnetic and transport properties. Our findings provide compelling evidence that magnetism can be engineered through DOS control at a phase crossover, establishing a pathway for the rational design of tunable quantum materials.

\end{abstract}

\red{\section{Keywords} 
Oxide ultra-thin film, Van Hove singularity, Quantum well state, ARPES, DFT+DMFT}
\section{Introduction}

The presence of a high density of states (DOS) has garnered significant attention as a key ingredient for discovering and manipulating exotic material properties. The high DOS can arise from the intrinsic structure or dimensionality of a material, with prominent examples including the emergence of flat bands~\cite{cho2021emergence,kang2022twofold,luo2023unique}, two-dimensional (2D) van Hove singularity (VHS)~\cite{PhysRev.89.1189,steppke2017strong,jerzembeck2023upper,mori2019controlling}, and quantum confinement effects~\cite{sakuragi2014thickness,kobayashi2015origin,sakuragi2018spontaneous}. When this high DOS is located near the Fermi level ($E_F$), it enhances the electron-electron interactions, leading to increased susceptibility to electronic instabilities. Consequently, these instabilities can give rise to a variety of correlated phases, including unconventional superconductivity~\cite{PhysRevLett.87.187004,piriou2011first,hu2022rich}, magnetic order~\cite{tserkovnyak2005nonlocal,goodenough1967narrow,liu2019magnetism}, \red{electronic nematicity~\cite{marshall2017growth,marshall2018electron},} charge-density waves~\cite{cho2021emergence,lin2021complex,luckin2024controlling}, and itinerant ferromagnetism~\cite{zong2023inducing,knuppel2025correlated,wahle1998microscopic}, highlighting the critical role of a high DOS near $E_F$ in driving collective phenomena.

Among these instabilities, itinerant ferromagnetism provides a compelling framework for directly linking a high DOS to an ordered state through the Stoner model. According to the Stoner model, ferromagnetism emerges when the product of the effective electron-electron interaction strength ($I$) and the nonmagnetic DOS at $E_{\rm F}$ ($N_0$) satisfies the condition $IN_0$~$\geq$~1~\cite{stoner1938collective}. Controlling a high DOS near $E_{\rm F}$ has been explored as a means to induce ferromagnetic states in certain systems. For instance, reducing dimensionality can generate a sharp increase in the DOS via the formation of a 2D VHS, which in turn stabilizes ferromagnetism~\cite{zong2023inducing}. Similarly, aligning a VHS at $E_{\rm F}$ through external fields has also been shown to induce itinerant ferromagnetism~\cite{knuppel2025correlated}. Despite these advances, practical applications remain limited by the low Curie temperature $T_{\rm C}$ (approximately 3~K) and fabrication challenges. Moreover, the microscopic origin of ferromagnetism stabilized by such a high DOS near $E_{\rm F}$ remains incompletely understood, particularly regarding the nature of spin-split electronic bands.

\begin{figure}
	\includegraphics[width=1\textwidth]{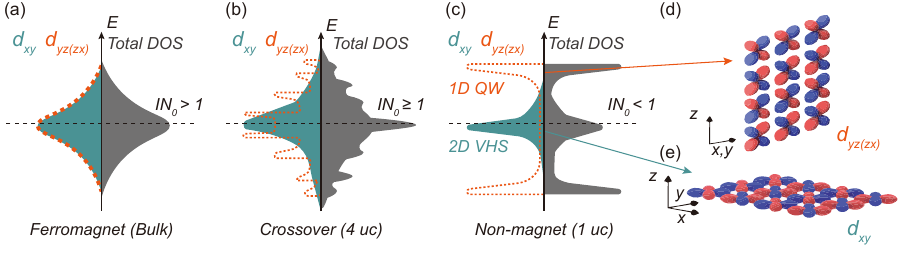} 
\caption{{\bf Evolution of the Ru $t_{2g}$ orbital density of states (DOS) with film thickness.} 
(a–c) Schematic DOS for (a) ferromagnetic bulk SrRuO$_3$ (SRO), (b) a 4 unit-cell (uc) SRO film at the magnetic crossover, and (c) a non-magnetic 1 uc SRO film, shown without considering electron correlations. 
(d,e) $d_{yz(zx)}$ and $d_{xy}$ orbitals, which give rise to one-dimensional quantum well (1D QW) states and a two-dimensional van Hove singularity (2D VHS) in ultrathin SRO films, respectively. 
} 

	\label{fig:1}
\end{figure}

To address these challenges, a 4~unit-cell (uc) SrRuO$_3$ (SRO) film provides a unique platform for investigating the controllability of a high DOS near $E_{\rm F}$~\cite{han2016ferromagnetism, kim2023electric, chang2009fundamental}. In bulk SRO, three Ru $t_{2g}$ orbitals (Ru 4$d_{xy}$, 4$d_{yz}$, and 4$d_{zx}$) contribute to multiple VHSs near $E_{\rm F}$, making it difficult to isolate the effect of each orbital (Fig. 1(a)). By contrast, quantum confinement in a 1~uc SRO film lifts the $t_{2g}$ orbital degeneracy. Consequently, only the two-dimensional (2D) VHS from the $d_{xy}$ orbital contributes significantly to the DOS at $E_{\rm F}$, while the $d_{yz/zx}$ orbitals form discrete quantum well (QW) states away from $E_{\rm F}$ (Fig. 1(c))~\cite{sohn2021observation,chang2009fundamental}. As a result, the limited DOS from this single VHS is insufficient to satisfy the Stoner criterion ($IN_0$~$\geq$~1). The limited DOS, together with the intrinsic instability of long-range magnetic order in two-dimensional systems described by the Mermin–Wagner theorem~\cite{mermin1966absence}, results in a non-magnetic ground state~\cite{kim2022tunable}.

The situation changes in the 4~uc SRO film (Fig. 1(b)), where inter-layer tunneling of the $d_{yz/zx}$ orbitals introduces additional one-dimensional (1D) QW states near $E_{\rm F}$ (Fig. 1(d)). These 1D QW \red{states}, together with the existing 2D VHS from the $d_{xy}$ orbital (Fig. 1(e)), collectively enhance the DOS at the Fermi level, $N_0$, thereby satisfying the Stoner criterion and stabilizing a ferromagnetic ground state. This mechanism is well reproduced by our density functional theory + dynamic mean-field theory (DFT+DMFT) calculations (see Section \uppercase\expandafter{\romannumeral1} of the Supplementary Materials (SM). Therefore, a 4 uc SRO film is situated precisely at the crossover from the non-magnetic (1~uc) to the ferromagnetic (bulk) phase, exhibiting high tunability under electron doping.

In this work, we demonstrate the direct control of emergent ferromagnetism driven by various dimensional quantum points in 4~uc SRO. Two complementary experimental approaches are employed: angle-resolved and spin-resolved photoemission spectroscopy (ARPES and SRPES) with alkali metal dosing (AMD), and transport measurements using the ionic-liquid gating method. Our density functional theory combined with dynamic\red{al} mean-field theory (DFT+DMFT) calculations provides strong theoretical support for the experimental observations.

\section{Results and discussion}
\begin{figure*}
\includegraphics[width=1\textwidth]{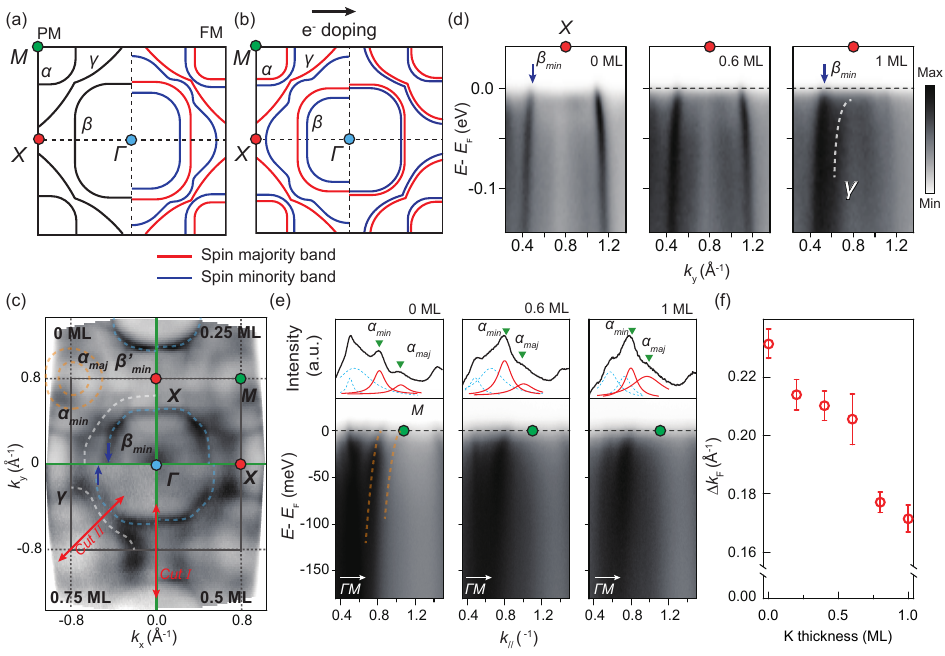}
\vspace{-0.5cm}
\caption{{\bf Angle-resolved photoemission spectroscopy (ARPES) results of a 4~uc SRO thin film.} 
(a) Schematic of the paramagnetic (PM) and ferromagnetic (FM) Fermi surfaces. Red, blue, and green circles correspond to the high symmetry $X$, $\Gamma$, and $M$ points, respectively.
(b) Schematic of Fermi surfaces with electron doping. The spin-split gap between the spin-majority and spin-minority bands decreases with electron doping~\cite{sohn2021sign}. 
(c) Fermi surfaces of the 4~uc SRO as a function of K dosing coverage. Each Fermi surface is obtained by integrating over an energy window of $E_{\rm F}$~$\pm$~10~meV, where $E_{\rm F}$ is the Fermi level. ML indicates a monolayer (see Fig. 3(d) for the definition of ML). The $\gamma$ bands are indicated by white dotted lines. The spin-majority and -minority $\alpha$ bands ($\alpha_{maj}$ and $\alpha_{min}$) are indicated by orange dotted lines, while the spin-minority $\beta$ and its folded bands ($\beta_{min}$ and $\beta'_{min}$) are indicated by blue dotted lines. The navy-colored arrows show that the Fermi momentum, {\it k$_F$}, of $\beta_{min}$ band increases with K dosing. 
(d,e) High-symmetry cuts near $X$ (Cut I) and $M$ (Cut II) with 0 ML (left), 0.6 ML (middle), and 1 ML (right) K coverage, respectively.
The upper panel of (e) shows MDCs with fitting results at $E_{\rm F}$. The green inverted triangles indicate the fitted peak positions of the $\alpha_{min}$ and $\alpha_{maj}$ peaks. 
(f) $\Delta$$k_{\rm F}$ as a function of K thickness. $\Delta k_{\rm F}$ represents the difference in $k_{\rm F}$ between $\alpha_{min}$ and $\alpha_{maj}$. All data were obtained at 6~K.}

\label{fig:3}
\end{figure*}

We control the electronic structure of a 4~uc SRO thin film using AMD. The film is grown by pulsed laser deposition (see Section \uppercase\expandafter{\romannumeral2} in SM for details of the growth process), and potassium (K) atoms serve as \red{electron-donor dopants.} Figure 2(a) presents schematic Fermi surfaces for the 4~uc SRO film in both the paramagnetic and the ferromagnetic states. In the paramagnetic state (left panel), three Ru 4$d$ $t_{2g}$ bands—$\alpha$, $\beta$, and $\gamma$—cross $E_{\rm F}$~\cite{sohn2021sign,shai2013quasiparticle,sohn2021observation}. In the ferromagnetic state (right panel), these bands split into spin-majority and spin-minority bands~\cite{hahn2021observation}, resulting in six bands crossing $E_{\rm F}$.

To investigate the evolution of the electronic structure with electron doping (Fig. 2(b)), we conduct ARPES measurements on the 4~uc SRO at 6~K with K dosing. Figure 2(c) shows the Fermi surface maps of SRO for different K coverage levels: 0, 0.25, 0.5, and 0.75 monolayers (ML), respectively. \red{Here, we quantify the K coverage in monolayers (ML), where 1 ML is defined as the coverage at which the spin polarization as a function of K dosing (discussed below) begins to saturate (Fig. 3(d)).} The $\gamma$ band near the $X$ point is marked by white dashed lines in the Fermi surface map. At 0 ML K coverage, the $\gamma$ band exhibits an electron-like band, but progressively transforms into a hole-like band at 0.75 ML K coverage, as previously reported (Fig. 2(c) and (d))~\cite{kim2023electric}. This change is in line with a shift of VHS across the Fermi level. \red{Note that in our ARPES measurements performed with linearly vertical (LV) polarized He-I$\alpha$ photons ($h\nu$ = 21.2 eV), the $\gamma$ band at 0~ML along $k_x = 0$ is strongly suppressed due to matrix-element effects, making it difficult to distinguish~\cite{iwasawa2012high,iwasawa2020high}.} 

The spin-majority and spin-minority $\alpha$ bands, $\alpha_{maj}$ and $\alpha_{min}$, which encompass the $M$ point, are marked by orange dotted lines (Fig. 2(c), top-left). The spin-minority $\beta$ band and its folded counterpart, $\beta_{min}$ and $\beta_{min}'$, respectively, are indicated by blue dotted lines, where $\beta_{min}'$ corresponds to the folded band in the second Brillouin zone. Navy arrows in the top-left and bottom-left panels of Fig. 2(c) indicate the Fermi momentum ($k_{\rm F}$) of $\beta_{min}$ along the $k_x$~=~0 direction for 0~ML and 0.75~ML K-dosed SRO, respectively. These arrows clearly show that the $\beta_{min}$ expands as the K coverage increases.

We next focus on the spin-split bands arising from ferromagnetism. Figure 2(e) shows the high-symmetry cut\red{s} along $\Gamma$-$M$ (Cut II) and the corresponding MDCs at $E_{\rm F}$. The $\alpha_{maj}$ and $\alpha_{min}$ bands are clearly observed~\cite{sohn2021sign}. To examine changes in band splitting, we fit each MDC with Lorentzian functions. The extracted peak positions are marked as green inverted triangles in the upper panel of each figure. To track the evolution of the peak distance between $\alpha_{maj}$ and $\alpha_{min}$, we plot $\Delta k_F$, the distance between the two peaks, as a function of K coverage (Fig. 2(f)). $\Delta k_F$ is 0.23~\AA$^{-1}$ in pristine SRO and decreases to 0.17~\AA$^{-1}$ with 1.0~ML of K, indicating a reduction in spin splitting with electron doping (see Section~\uppercase\expandafter{\romannumeral4} in SM for the position of each $\alpha$ band). 

Our high-resolution ARPES results reveal that K doping leads to two key changes: (i) a shift of the VHS away from the Fermi level (Fig. 2(d)), and (ii) the peak distance between $\alpha_{maj}$ and $\alpha_{min}$ decreases (Fig. 2(f)). These observations indicate K dosing results in an increase in $E_{\rm F}$ and a corresponding reduction in spin splitting. Taken together, these results suggest that ferromagnetism is progressively suppressed as $E_{\rm F}$ increases.

According to the Stoner criterion, the spin polarization, $P = (N_{\uparrow}-N_{\downarrow})/N$, where $N_{\uparrow}$ ($N_{\downarrow}$) are the numbers of spin-majority and spin-minority electrons, respectively, and $N$ is the total number of electrons, is expected to decrease as ferromagnetism is suppressed. To investigate this, we perform SRPES to track the evolution of spin polarization as a function of K coverage. 
\begin{figure}
\includegraphics[width=0.48\textwidth]{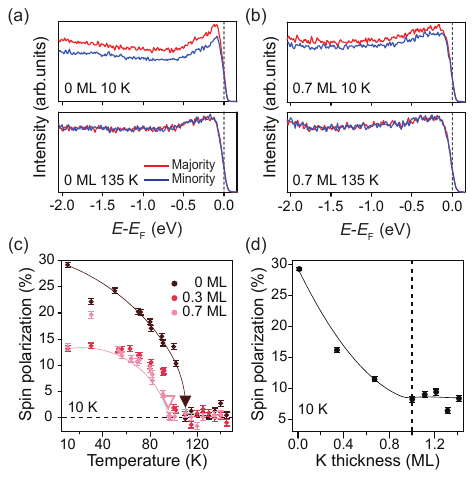}
 \centering
\caption{{\bf Spin-resolved photoemission spectroscopy (SRPES) results of a 4 uc SRO thin film.}
(a,b) SRPES energy distribution curves for (a) 0~ML and (b) 0.7~ML K-dosed films measured at 10~K and 135~K. Spin polarization vanishes at 135~K. All measurements were taken at the $\Gamma$ point. 
(c) Temperature-dependent spin polarization measurements at each dosing step. The temperature at which polarization reaches 0~$\%$ is defined as $T_{\rm C}$. The filled triangle indicates the $T_{\rm C}$ of pristine 4~uc SRO, while the open triangle represents the reduced $T_{\rm C}$ for 0.7~ML K coverage.
(d) Spin polarization as a function of K coverage. Here, 1~ML of K is defined as the point where the spin polarization begins to saturate, indicated by a black dotted line. The solid lines in (c) and (d) serve as guides to the eye for the temperature- and K-thickness-dependent changes in spin polarization, respectively. \red{Error bars represent the experimental uncertainty of the SRPES measurements (see SM section \uppercase\expandafter{\romannumeral3} for the detail).}}
\label{fig:4}
\end{figure}
Figure 3(a) shows spin-resolved energy distribution curves (EDCs) of pristine 4~uc SRO measured at 10~K and 135~K. A clear intensity difference between majority and minority photoelectrons is observed at 10~K, while no such difference is seen at 135~K, indicating the loss of spin polarization above $T_{\rm C}$. Figure 3(b) presents spin-resolved EDCs of 0.7~ML K-dosed SRO measured at the same temperatures. At 10~K, the intensity difference is reduced relative to the pristine sample, consistent with the suppression of ferromagnetism. At 135~K, the intensity difference vanishes in both cases.

To investigate the evolution of spin polarization in greater detail, we perform SRPES measurements with fine temperature steps. Figure 3(c) shows the temperature-dependent spin polarization as a function of K coverage. Here, the spin polarization, $P$, is defined as $(I_{\rm maj}-I_{\rm min})/(I_{\rm maj}+I_{\rm min})$, where $I_{\rm maj}$ and $I_{\rm min}$ are the intensities of the spin-majority and spin-minority photoelectrons at the $\Gamma$ point. The intensity is integrated from $E_{\rm F}$~-~2~eV to $E_{\rm F}$ in each EDC (see section~\uppercase\expandafter{\romannumeral3} in SM). At low temperatures, the spin polarization is nonzero; however, it gradually decreases with increasing temperature and eventually saturates at zero above $T_{\rm C}$ marked by inverted triangles~\cite{fujiwara2018origins,pickel2010magnetic}. Notably, $T_{\rm C}$ decreases from 110~K to 96~K with increasing K coverage. Figure 3(d) also shows that the spin polarization at 10~K decreases with K dosing. In the pristine film, the spin polarization is 30~\%, and it decreases monotonically with K dosing, eventually saturating at approximately 8.8~\%. \red{This systematic suppression of spin polarization with K dosing is not unique to the 4~uc film. A similar tendency is observed in the 3~uc SRO film (see Fig. S6 in SM for SRPES data of the 1~uc and 3~uc SRO films).}

We obtain consistent results from SRPES and ARPES measurements. SRPES reveals that with K dosing: (i) $T_{\rm C}$ decreases, and (ii) spin polarization below $T_{\rm C}$ is also reduced. Similarly, ARPES shows a decrease in the spin-split gap between $\alpha_{maj}$ and $\alpha_{min}$ with increasing K dosing (see Section \uppercase\expandafter{\romannumeral4} of the SM for the detailed analysis). Together, these results indicate that ferromagnetism is progressively suppressed as $E_{\rm F}$ increases.

Changes in magnetic properties are generally accompanied by corresponding variations in transport behavior~\cite{ye2018massive,kim2018large,sohn2021sign}. To complement the photoemission results, we perform transport measurements on 4~uc~SRO films capped with 10~uc~SrTiO$_3$ layers using the ionic-liquid gating method. The STO capping layer prevents chemical reactions between SRO and the ionic liquid~\cite{bisri2017endeavor,li2020reversible}. The ionic-liquid gating method is a well-established technique for tuning $E_{F}$ via gate voltage ($V_{\rm G}$) control~\cite{nakano2012collective,kim2021capping,lin2021electric}. When a positive (negative) $V_{\rm G}$ is applied, cations (anions) in the ionic liquid migrate toward the film, resulting in electron (hole) doping. 
\begin{figure}
\includegraphics[width=0.48\textwidth]{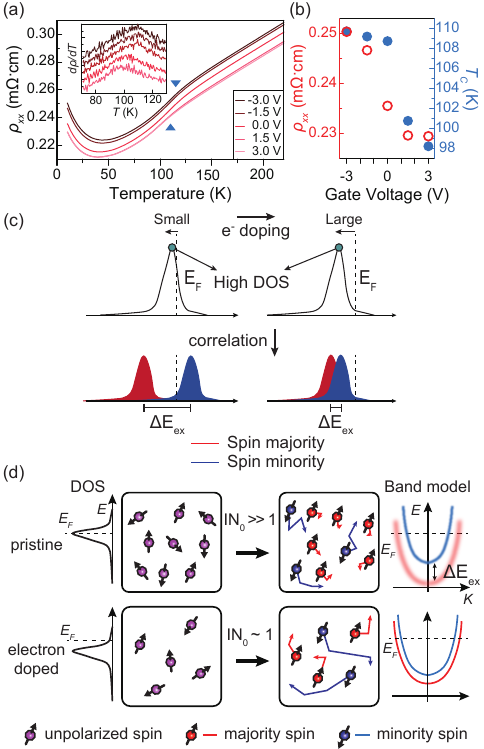}
\vspace{0.5cm}
 \centering
\caption{{\bf Control of transport properties of SRO via the ionic-liquid gating method.} 
(a) Gate-voltage-dependent resistivity, $\rho_{xx}$, of SRO as a function of temperature. The gate voltage is applied from -3~V to 3~V. Blue triangles mark the kink in the curves, corresponding to the Curie temperature, $T_{\rm C}$. (Inset) Derivative of resistivity near the kink. 
(b) Gate-voltage-dependent $\rho_{xx}$ at 10~K and $T_{\rm C}$.
(c) Schematic\red{s of a high DOS} at different Fermi levels ($E_{\rm F}$), with the corresponding spin-resolved DOS. 
The high DOS peak positions for non-magnetic and spin-polarized SRO are marked by a green dot. $\Delta E_{ex}$ denotes the ferromagnetic exchange energy.
(d) Schematic illustration of the DOS, electrons, and electronic structure in SRO. All electrons in the schematic are positioned at $E_{\rm F}$. The motion of electrons depends on the relative alignment between the high DOS position and $E_{\rm F}$, leading to changes in both ferromagnetism and conductivity of SRO.}
\label{fig:5}
\end{figure} 

Figure 4(a) shows the longitudinal resistivity $\rho_{xx}$ measured under different gate voltages $V_{\rm G}$. A kink in $\rho_{xx}$, corresponding to $T_{\rm C}$~\cite{allen1996transport}, is marked by blue triangles. As $V_{\rm G}$ increases from -~3~V to 3~V, both $T_{\rm C}$ and $\rho_{xx}$ decrease. The derivative of resistivity, d$\rho$/d$T$, plotted in the inset, exhibits a peak at the kink position, illustrating the evolution of $T_{\rm C}$ as a function of $V_{\rm G}$. This analysis further confirms that $T_{\rm C}$ decreases with increasing $V_{\rm G}$. 

Figure 4(b) presents the dependence of both $T_{\rm C}$ and $\rho_{xx}$ at 10~K on $V_{\rm G}$. Consistently, both quantities decrease as $V_{\rm G}$ increases \red{(see Section \uppercase\expandafter{\romannumeral5} in SM for gate voltage dependency in the Hall effect measurements).} The trend in $T_{\rm C}$ aligns with our SRPES results, in which both $T_{\rm C}$ and spin polarization decrease with K dosing. Interestingly, the reduction in $\rho_{xx}$ with increasing $V_{\rm G}$ suggests that the metallicity of SRO is enhanced as ferromagnetism is weakened.

Our findings from ARPES, SRPES, transport measurements, and DFT+DMFT are schematically summarized in Fig.~4(c) and 4(d). With electron doping, the relative alignment between the high-DOS point and $E_F$ shifts, thereby modifying the DOS at $E_{\rm F}$. This change weakens the electron-electron interactions through two complementary mechanism\red{s}: (i) at low energies, the reduced $N(E_{F})$ decreases the available phase space for electron-electron scattering, and (ii) at high energies, electron doping changes the nominal $t_{2g}$ occupancy (N) from $N\approx4$~(Ru$^{4+}$) towards N~=~5. This occupancy change intrinsically weakens correlation effects, which are strongly enhanced in the $N\approx4$~(Ru$^{4+}$) configuration due to Hund's coupling~\cite{georges2013strong}. The overall reduction of interactions suppresses ferromagnetism and lowers the ferromagnetic exchange energy, $\Delta E_{ex}$. Simultaneously, the decrease in the number of electrons participating in scattering reduces the scattering rate of itinerant electrons. Consequently, metallicity is enhanced with electron doping. These conclusions are supported by the following experimental observations;
\begin{table}[htbp]
\centering
\renewcommand{\arraystretch}{1.2}  
\begin{tabular}{ll}
(i) & reduction in spin splitting between $\alpha_{\text{maj}}$ and $\alpha_{\text{min}}$ (ARPES); \\
(ii) & decrease in spin polarization (SRPES); \\
(iii) & lowering of $T_{\rm C}$ (SRPES and transport); \\
(iv) & decrease in resistivity under ionic-liquid gating (transport).\\
\end{tabular}
\vspace{1ex}
\end{table}

In summary, we demonstrate that ferromagnetism in ultrathin SRO films can be effectively controlled by tuning the relative position of $E_{\rm F}$ with respect to the high DOS point. Using ARPES, we reveal the evolution of the electronic structure and spin-split bands as $E_{\rm F}$ shifts, while SRPES confirms that electron doping suppresses ferromagnetism by reducing both $T_{\rm C}$ and spin polarization. Complementary transport measurements show a concurrent enhancement of metallicity with electron doping. These experimental observations are supported by DFT+DMFT calculations, which highlight the interplay between electron correlations, metallicity, and ferromagnetism in SRO. Our results underscore the critical role of the high DOS such as 1D QW and 2D VHS in modulating electronic and magnetic properties, offering a promising platform for engineering quantum phases and advancing spintronic applications through controlled doping.

\red{\section{Supporting Information}
Calculation and experimental methods, additional experimental data for SRPES, ARPES, and Ionic-liquid gating.
}
\begin{acknowledgement}

This work was supported by the National Research Foundation of Korea (NRF) grant funded by the Korea government (MSIT) (No. 2022R1A3B1077234) and the Global Research Development Center (GRDC) Cooperative Hub Program through the NRF funded by the MSIT (Grant No. RS-2023-00258359). This work was also supported by the Institute of Applied Physics, Seoul National University. This research was supported by Basic Science Research Program through the National Research Foundation of Korea(NRF) funded by the Ministry of Education (No. RS-2019-NR040081). This work was supported by the National Research Foundation of Korea (NRF) grant funded by the Korea government (MSIT) (Grant No. RS-2025-00513951). MK was supported by Korea Institute for Advanced Study (KIAS) individual Grants (No. CG083502). The DFT+DMFT calculation is supported by the Center for Advanced Computation at KIAS. 

\end{acknowledgement}


\providecommand{\latin}[1]{#1}
\makeatletter
\providecommand{\doi}
  {\begingroup\let\do\@makeother\dospecials
  \catcode`\{=1 \catcode`\}=2 \doi@aux}
\providecommand{\doi@aux}[1]{\endgroup\texttt{#1}}
\makeatother
\providecommand*\mcitethebibliography{\thebibliography}
\csname @ifundefined\endcsname{endmcitethebibliography}  {\let\endmcitethebibliography\endthebibliography}{}

\end{document}